\documentclass[aip,pop,reprint,amsmath,amssymb]{revtex4-2}

\usepackage{graphicx}
\usepackage{dcolumn}
\usepackage{bm}
\usepackage{etoolbox}
\usepackage{hyperref}
\usepackage{subcaption}

\makeatletter
\def\@citess#1{\leavevmode\unskip\penalty\@M\ [\@cite{#1}{}]}
\makeatother

\newcommand{\sechead}[1]{\par\medskip\noindent\emph{#1.}---}
\newcommand{\He}{\mathrm{He}}

\makeatletter
\def\@email#1#2{%
 \endgroup
 \patchcmd{\titleblock@produce}
  {\frontmatter@RRAPformat}
  {\frontmatter@RRAPformat{\produce@RRAP{*#1\href{mailto:#2}{#2}}}\frontmatter@RRAPformat}
  {}{}
}%
\makeatother

\begin{document}

\title{Statistical structure of canonical kinetic equilibrium with sheared flow}

\author{Daniel W. Crews}
\email{daniel.crews@zap.energy}
\affiliation{Zap Energy, Everett, WA 98203, USA}

\date{\today}

\begin{abstract}
In isothermal kinetic equilibrium of a plasma with sheared flow,
the electrostatic potential is the cumulant generating function of the
velocity distribution, encoding all non-Maxwellian statistics.
Consequently, of the polynomial-form flux-function equilibrium
flows, only the linear one is admissible, and every other
equilibrium sheared flow has non-zero cumulants to all orders.
The identity is shown for a sheared-flow screw pinch and
specialized to the Z pinch, theta pinch, and Harris sheet.
A simple sheared-flow Bennett pinch illustrates the principal results, namely
weaker magnetization yields more non-Maxwellian features, 
and co- and counter-current flow shear have distinct statistics.
The identity also yields a method for initializing kinetic simulations.
\end{abstract}

\maketitle

The canonical kinetic equilibrium distribution function $f \propto F(P)\,e^{-\beta H}$
with $F$ a function of the canonical momenta,~\cite{gratreau1978} $H$ the single-particle Hamiltonian,
and $\beta = 1/kT$ constant, underlies a long lineage of
magnetic equilibria from sheared magnetic fields,~\cite{channell1976}
current sheets,~\cite{harris1962, mahajan1989, mahajan2000sheared}
force-free plasmas,~\cite{janaki2012, wu2024}
and recent two-species~\cite{cordonnier2022, cordonnier2025} and hybrid~\cite{kaltsas2025hybrid}
screw-pinch constructions.
In simple cases these equilibrium distributions are drifting Maxwellians,
whose perturbations and stability are well studied.~\cite{bott2024}
More generally, however, the kinetic equilibrium distribution
is non-Maxwellian.~\cite{ewart2025}

Equilibrium non-Maxwellian features such as pressure anisotropy are well known to occur
with sheared flows.~\cite{delsarto2016, delsarto2017, mahajan2000sheared}
These canonical distributions have a well-developed connection to Hermite
expansions and the Weierstrass transform.~\cite{suzuki2008novel, allanson2016}
This letter examines sheared-flow equilibria 
and emphasizes a structural identity, namely that the electrostatic potential of the sheared flows, 
expressed as a flux function, is the cumulant generating function of the equilibrium distribution. 
It follows that all the non-Maxwellian statistics of isothermal flow equilibria are encoded in the
electric potential of the sheared flow and its successive derivatives.
The higher cumulants of the equilibria (and so the moments) obey a power law,
as shown in this letter for the Bennett Z pinch.

The identity applies to both electron and ion species, but this letter focuses on
the ions and typically drops species subscripts, as the effect depends on Larmor radius
and the electrons thus remain more effectively Maxwellian.
The identity applies directly to axisymmetric configurations with
translational symmetry, exemplified by present sheared-flow Z pinches,~\cite{shumlak2020}
and by space plasma current sheets where power-law scaling of high-order moments
has been observed in Magnetospheric Multiscale (MMS) Mission data.~\cite{abid2024, richard2025}
Analogous non-Maxwellian features arise in anisotropic mirrors,~\cite{frank2026}
centrifugal mirrors,~\cite{stoltzfusdueck2025,ball2025cmfx,hazeltine2026}
levitated dipoles,~\cite{chisholm2025} and stellarators.~\cite{helander2014}

% collisional Chapman--Enskog distortions~\cite{bott2024}

\sechead{Kinetic equilibrium and electric field}Consider an axisymmetric
cylindrical plasma with magnetic field
$\vec B = B_\theta(r)\hat\theta + B_z(r)\hat z$ generated by a
vector potential $\vec A = A_\theta(r)\hat\theta + A_z(r)\hat z$,
so that $B_\theta = -dA_z/dr$ and $B_z = r^{-1}d\psi/dr$ with
$\psi(r) \equiv rA_\theta(r)$ the poloidal flux per radian, together
with an electrostatic potential $\varphi(r)$.
The single-particle constants of motion are
\begin{align}
  P_z &= m v_z + q A_z, \\
  P_\theta &= m r v_\theta + q\psi, \\
  H &= \tfrac{1}{2} m v^2 + q\varphi,
  \label{eq:invariants}
\end{align}
and any distribution of the form $f = f(P_\theta, P_z, H)$
satisfies the steady Vlasov equation.~\cite{vogman2019}
Radial isothermality selects the canonical energy dependence
\begin{equation}
  f(r, \vec v) = F(P_z, P_\theta)\, e^{-\beta H},
  \label{eq:ansatz}
\end{equation}
with $\beta = 1/kT$ spatially uniform.~\cite{gratreau1978}
% The function $F$ carries all information about flow
% shear and electromagnetic structure.

Assuming ion--electron equilibrium, 
summing the species momentum balances 
$q_s n_s(\vec E + \vec v_s \times \vec B) = \nabla p_s$ 
and using $\nabla (p_e + p_i) = \vec{\jmath}\times\vec{B}$
eliminates the pressure gradient under $T_i = T_e$, $n_i = n_e$, and $Z_i = 1$, 
leaving $\vec E = -\tfrac{1}{2}(\vec v_i + \vec v_e)\times\vec B$.
Decomposing the species flows as
$\vec v_i = \vec u_0 + \vec v$ and $\vec v_e = -\vec u_0 + \vec v$
separates the counter-streaming drifts
$\vec u_0 = \tfrac{1}{2}(\vec v_i - \vec v_e)$ from the common flow
$\vec v = \tfrac{1}{2}(\vec v_i + \vec v_e)$ shared by both
species, the latter defined up to an additive constant;
fixing this constant is equivalent to a choice of frame.~\cite{yoon2024, crews2025transitional}
In this way one finds
\begin{equation}
  \vec E \;=\; -\,\vec v\times\vec B.
  \label{eq:Ohm}
\end{equation}
In the screw-pinch context, this letter uses the rigid drift
\begin{equation}
  2\vec u_0 \;=\; -\,r\,\Omega_0\,\hat\theta + u_{0z}\,\hat z,
  \label{eq:u0-form}
\end{equation}
with $\Omega_0$, $u_{0z}$ constants to isolate the effect of flow shear.

\sechead{Flow and potential as flux functions}Writing a flow component as a function of the corresponding flux
requires two conditions. First, the flux should be monotonic in $r$,
or equivalently the magnetic field component it sources does not change sign; 
and second, the flow as a function of the flux is analytic on a disk in the 
complex flux plane containing the range of fluxes. 
Both are satisfied for the examples considered later.

The radial part of Eq.~\eqref{eq:Ohm} is
$E_r = v_z B_\theta - v_\theta B_z$, and using $E_r = -d\varphi/dr$,
$B_\theta = -dA_z/dr$, and $B_z = r^{-1}d\psi/dr$ gives the potential form
\begin{equation}
  \frac{d\varphi}{dr} \;=\; v_z\,\frac{dA_z}{dr} + \Omega\,\frac{d\psi}{dr},
  \qquad \Omega \equiv v_\theta/r.
  \label{eq:phi-dr}
\end{equation}
Now, for $\varphi$ to exist as a single-valued potential of
the two flux coordinates $(A_z, \psi)$, the right-hand side of
Eq.~\eqref{eq:phi-dr} must be an exact differential, requiring for compatibility
\begin{equation}
  \left(\frac{\partial v_z}{\partial \psi}\right)_{A_z}
  \;=\;
  \left(\frac{\partial \Omega}{\partial A_z}\right)_\psi,
  \label{eq:compat}
\end{equation}
in which case $v_z = \partial\varphi/\partial A_z$ and
$\Omega = \partial\varphi/\partial\psi$.  Equation~\eqref{eq:compat}
is the closure condition $d\omega = 0$ for the 1-form
$\omega \equiv v_z\,dA_z + \Omega\,d\psi$ on flux-coordinate space;
the Poincar\'e lemma then guarantees $\omega = d\varphi$.~\cite{weintraub2014}

If each flow component is a flux function of its own magnetic
potential (\textit{i.e.}, $v_z = v_z(A_z)$ and $\Omega = \Omega(\psi)$)
then Eq.~\eqref{eq:compat} is satisfied and
Eq.~\eqref{eq:phi-dr} integrates to a separable potential
\begin{equation}
  \varphi(A_z, \psi) \;=\; \varphi_z(A_z) + \varphi_\theta(\psi),
  \label{eq:phi-split}
\end{equation}
with $d\varphi_z/dA_z = v_z$ and $d\varphi_\theta/d\psi = \Omega$.
Assuming a correspondingly separable momentum function
$F(P_z, P_\theta) = F_z(P_z)\,F_\theta(P_\theta)$ then factorizes
the kinetic distribution.

Sheared-flow screw-pinch equilibria are in general non-separable,
but this letter develops the simpler, separable case.
The simplest separable screw-pinch equilibrium
(no sheared flow, uniform axial and azimuthal drift)
is the recently constructed self-consistent solution of
Refs.~\citenum{cordonnier2022} and \citenum{cordonnier2025};
our framework supplies the equilibrium
statistics of sheared flows on such a separable equilibrium.

The Taylor expansions of the flow components are
\begin{align}
  v_z(A_z) &= \sum_{n=1}^{\infty} \frac{c_{n,z}}{n!}\, A_z^n,
  \label{eq:vz-series}\\
  \Omega(\psi) &= \sum_{n=1}^{\infty} \frac{c_{n,\theta}}{n!}\, \psi^n,
  \label{eq:Omega-series}
\end{align}
with $c_{n,z}$ and $c_{n,\theta}$ the shear coefficients,
and the constant rigid drifts $u_{0z}$ and $\Omega_0$
of Eq.~\eqref{eq:u0-form} are omitted.
Here the angular velocity $\Omega$, not the azimuthal velocity
$v_\theta$, is naturally the flux function.
Integration gives the potentials
\begin{align}
  \varphi_z(A_z) &= \sum_{n=1}^{\infty} \frac{c_{n,z}}{(n+1)!}\, A_z^{n+1},
  \label{eq:phiz-series}\\
  \varphi_\theta(\psi) &= \sum_{n=1}^{\infty} \frac{c_{n,\theta}}{(n+1)!}\, \psi^{n+1}.
\end{align}
% up to an additive gauge constant.  

\sechead{Gram--Charlier series}Within the separable ansatz, the axial and
azimuthal marginal distribution functions are analyzed separately. 
For the axial direction,
$f_z(v_z, r) \propto F_z(P_z)\, e^{-v_z^2/(2\sigma^2)}$; introduce
dimensionless variables $y \equiv v_z/\sigma$ and $a \equiv qA_z/(m\sigma)$, 
so that $P_z/(m\sigma) = y + a$, and write $G_z(s) \equiv F_z(m\sigma s)$.
The probabilist's Hermite expansion of $G_z(y+a)$ in $y$ reads
\begin{equation}
  G_z(y+a) = \sum_{n=0}^{\infty} \frac{g_z^{(n)}(a)}{n!}\, \He_n(y),
  \label{eq:hermite}
\end{equation}
with $g_z(a) \equiv (2\pi)^{-1/2}\int G_z(y+a)\, e^{-y^2/2}\, dy$
and primes denoting $a$-derivatives.~\cite{suzuki2008novel, allanson2016}

The first moment of Eq.~\eqref{eq:hermite} at radius $r$ is
$\langle y\rangle = g_z'/g_z = d\ln g_z/da$, so
$\langle v_z\rangle(r) = \sigma\,d\ln g_z/da$.
Under the rigid relative velocity assumption Eq.~\eqref{eq:u0-form},
the species flow is consistent with the common-flow,
$\sigma\,d\ln g_z/da = d\varphi_z/dA_z$, which integrates to
\begin{equation}
  g_z(a) \;=\; Z_z^{-1}\, e^{\Phi_z(a)},
  \qquad
  \Phi_z(a) \equiv \frac{q\,\varphi_z(A_z)}{m\sigma^2},
  \label{eq:g-Phi}
\end{equation}
identifying $g_z$ as the contribution to plasma density from $A_z$, 
with $Z_z$ the partition function.~\cite{mahajan1989}
Substituting Eq.~\eqref{eq:g-Phi} into Eq.~\eqref{eq:hermite} and using
Faa di Bruno's formula
$\partial_a^n e^{\Phi_z} = e^{\Phi_z}\,B_n(\Phi_z', \ldots, \Phi_z^{(n)})$,
with $B_n$ the complete Bell polynomials, gives the axial marginal distribution
as a Maxwellian modulated by a Gram--Charlier series,~\cite{blinnikov1998}
\begin{equation}
  f_z(v_z, r) \propto
    e^{-v_z^2/(2\sigma^2)}
    \sum_{n=0}^{\infty}
    \frac{B_n(\Phi_z',\ldots,\Phi_z^{(n)})}{n!}\,
    \He_n\!\left(\frac{v_z}{\sigma}\right).
  \label{eq:bell-form}
\end{equation}
The azimuthal marginal distribution follows by the same construction with
$\Phi_\theta(\psi) \equiv q\,\varphi_\theta(\psi)/(m\sigma^2)$ and
derivatives taken with respect to the flux $q\psi$.

\sechead{Statistics of sheared-flow equilibrium}For a probability density $p(x)$, the moment generating function (MGF)
$M(k) = \langle e^{kx}\rangle$ encodes the moments through its
Taylor expansion, and its logarithm
\begin{equation}\label{eq:cgf_def}
  \ln M(k) = \sum_{n=1}^{\infty} \frac{\kappa_n}{n!}\,k^n,
\end{equation}
by definition, generates the cumulants $\{\kappa_n\}$,
so $\ln M(k)$ is called the cumulant generating function (CGF).~\cite{vankampen2007}
Intuitively, the first cumulant $\kappa_1$ is the mean, $\kappa_2$ the variance,
$\kappa_3$ the skewness, and $\kappa_4$ the excess kurtosis.
Each cumulant is a polynomial in the moments which vanishes for a
Maxwellian beyond $n = 2$, so $\kappa_{n\geq 3} \neq 0$ measures
the non-Maxwellian part of the distribution function in a compact
form.~\cite{lukacs1970, blinnikov1998}

Applied to the marginal distribution of axial velocity at fixed
flux coordinate, the MGF of $y \equiv v_z/\sigma$ is
\begin{equation}
  M_y(k) \;=\;
  \frac{\int G_z(y+a)\,e^{ky - y^2/2}\,dy}
       {\int G_z(y+a)\,e^{-y^2/2}\,dy}.
  \label{eq:mgf-int}
\end{equation}
Completing the square via $ky - y^2/2 = -\tfrac{1}{2}(y-k)^2 + \tfrac{1}{2}k^2$
and shifting $y \to y+k$ in the numerator gives
\begin{equation}
  M_y(k) \;=\; e^{k^2/2}\,\frac{g_z(a+k)}{g_z(a)}.
  \label{eq:mgf-shift}
\end{equation}
Substituting Eq.~\eqref{eq:g-Phi},
\begin{equation}
  \ln M_y(k) \;=\; \frac{k^2}{2} + \Phi_z(a+k) - \Phi_z(a).
  \label{eq:lnpsi}
\end{equation}
The term $k^2/2$ is the generator of the Maxwellian thermal velocity. 
Equation~\eqref{eq:lnpsi} identifies $\Phi_z$
as the CGF of the non-Maxwellian features of the distribution.
Thus, \emph{the electric potential of a canonical kinetic equilibrium with sheared flow, 
expressed as a flux function, is itself the CGF of the distribution function,}
fully encoding the statistical structure of the equilibrium.
An analogous construction applied to the azimuthal motion $\Omega = v_\theta/r$
yields the same result, with the poloidal potential $\Phi_\theta(\psi)$
acting as the CGF of the $\Omega$-distribution.

Comparing the Taylor expansion of the shifted potential in
Eq.~\eqref{eq:lnpsi} with the CGF definition Eq.~\eqref{eq:cgf_def}
reads off the cumulants of $y = v_z/\sigma$ directly as successive
derivatives of $\Phi_z$ at the dimensionless axial flux $a$,
\begin{equation}
  \kappa_n^{v_z/\sigma}(a) \;=\; \Phi_z^{(n)}(a),
  \qquad n \geq 3,
  \label{eq:cumulants-axial}
\end{equation}
with mean $\kappa_1^{v_z/\sigma} = \Phi_z'(a) = v_z/\sigma$ as the flow 
and variance $\kappa_2^{v_z/\sigma} = 1 + \Phi_z''(a)$, \textit{i.e.}, 
the base temperature plus thermal anisotropy induced by the shear flow. 

An identical construction for the marginal distribution of azimuthal 
velocities, using dimensionless flux $b \equiv q\psi/(mr\sigma)$
and $\Phi_\theta(b) \equiv q\varphi_\theta/(m\sigma^2)$, gives
\begin{equation}
  \kappa_n^{v_\theta/\sigma}(b) \;=\; \Phi_\theta^{(n)}(b),
  \qquad n \geq 3.
  \label{eq:cumulants-azim}
\end{equation}
Equations \eqref{eq:cumulants-axial}--\eqref{eq:cumulants-azim} are
the central result of this letter: 
in a canonical sheared-flow kinetic equilibrium,
the statistical structure of the distribution function is encoded 
in suitably normalized derivatives of the electric potential with 
respect to the corresponding flux coordinate.
Equivalently, exponentiating the potential (recall $g_z = Z_z^{-1}\exp(\Phi_z)$)
turns the CGF into the MGF. The plasma density in flux
coordinates is therefore the moment generating function of the
distribution function.

\emph{Non-existence of polynomial flux-function equilibria.} 
A classical theorem forbids CGFs from polynomial form of
degree greater than two,~\cite{marcinkiewicz1939, lukacs1970}
as higher degree corresponds to distributions
with negative regions, physically inadmissible.
Because $d\varphi/dA = v$ raises the polynomial degree by one, a polynomial
flux-function flow of degree $N$ produces a $\Phi$ of degree $N+1$,
and the theorem thereby forces $N \leq 1$. 
Linear flux-function flow is therefore the unique admissible 
polynomial-form equilibrium, illustrated in the next section 
by canonical Z-pinch and theta-pinch examples. 
All other equilibrium flows have $\kappa_{n\geq 3} \neq 0$, meaning non-Maxwellian statistics.

\sechead{Linear flux-function flows}The simplest shear flow is a linear function of the magnetic flux,
for which Eq.~\eqref{eq:cumulants-axial} terminates the cumulant
hierarchy at $n = 2$.  Such equilibria support anisotropic
Maxwellians with shear-induced temperature variation along the
flow directions.

On the separable screw-pinch solution of Ref.~\citenum{cordonnier2022},
featuring uniform axial drift, rigid azimuthal rotation, and magnetic fluxes $A_z(r)$ and $\psi(r)$,
one may add linear flux-function flow shear in the axial and azimuthal directions,
$v_z(A_z) = -\alpha\,qA_z/m$ and $\Omega(\psi) = -\gamma\,q\psi/m$ to
induce temperature anisotropies
\begin{equation}
  \frac{T - T_{zz}}{T} \;=\; \alpha,
  \qquad
  \frac{T - T_{\theta\theta}}{T} \;=\; \gamma\,r^2.
  \label{eq:linear-anis}
\end{equation}
Two classical configurations are limits of this solution, both
solving the Liouville--Grad-Shafranov equation, with characteristic radii
set by self-consistency:

\emph{Z-pinch limit.}  With zero external field and no rotation,
the solution reduces to the Bennett equilibrium with flux 
$A_z(r) = (\mu_0 I_\infty/4\pi)\ln(1 + (r/r_p)^2)$ and density
$n(r) = n_0/(1 + (r/r_p)^2)^2$, where axial current $j_z$ generates
its own confining $B_\theta$.~\cite{bennett1934}
Axial flow shear as a linear flux function yields a radially uniform
flow-parallel anisotropy.
In the 1D limit (Harris current sheet, $\vec A = A_z(y)\hat z$),~\cite{harris1962, ceccherini2005} 
the construction reduces verbatim to $d\varphi/dA_z = v_z(A_z)$ and
recovers the shear-flow Harris sheet of Ref.~\citenum{mahajan2000sheared}.

\emph{Theta-pinch limit.} With zero axial current but
external field $B_{z,0}\hat z$, the coupled ODEs of Ref.~\citenum{cordonnier2022} 
reduce to Liouville's equation in $\psi$ with the exact solution
$\psi(r) = \tfrac{1}{2}B_z(0)\,r^2 + (2kT/e|\Omega_0|)\ln\cosh((r/r_c)^2)$
and density $n(r) = n_0\,\mathrm{sech}^2((r/r_c)^2)$.
The field $B_{z,0}$ at infinity is diamagnetically depressed to
$B_z(0) = (m_i+m_e)|\Omega_0|/(2e)$ on-axis, with $\Omega_0$ the
species drift angular velocity.
Sheared azimuthal rotation as a linear flux function yields the $T_{\theta\theta}$
shear-induced anisotropy of Eq.~\eqref{eq:linear-anis}.

\sechead{Non-Maxwellian equilibrium flows}On a Bennett Z-pinch equilibrium,
the flow $v_z = V_0\,(r/r_p)^2$ illustrates well the distribution function of general flows.
Inverting $A_z = A_0 \ln(1 + (r/r_p)^2)$ gives $r^2/r_p^2 = e^{-\lambda a} - 1$ 
with $\lambda \equiv m\sigma/(qA_0)$, yielding an exponential
flux function,
\begin{equation}
  v_z(a) = V_0\bigl(e^{-\lambda a} - 1\bigr)
         = V_0 \sum_{n=1}^{\infty} \frac{(-\lambda a)^n}{n!}.
  \label{eq:vz-pois}
\end{equation}
Integration yields the normalized electric potential
\begin{equation}
  \Phi_z(a) = \mu\Bigl[\frac{1 - e^{-\lambda a}}{\lambda} - a\Bigr],
  \qquad \mu \equiv V_0/\sigma,
  \label{eq:Phi-pois}
\end{equation}
from which proceed, using Eq.~\eqref{eq:cumulants-axial}, the cumulants
\begin{equation}
  \kappa_n^{v_z} = \sigma^n\,\mu\,(-\lambda)^{n-1}\,e^{-\lambda a(r)}
  \qquad (n \geq 3),
  \label{eq:cum-pois}
\end{equation}
with $\kappa_2^{v_z} = \sigma^2\bigl[1 - \mu\lambda\,e^{-\lambda a}\bigr]$.
% and $\kappa_1^{v_z} = V_0 r^2/r_p^2$.

\begin{figure}
  \centering
  \includegraphics[width=\columnwidth]{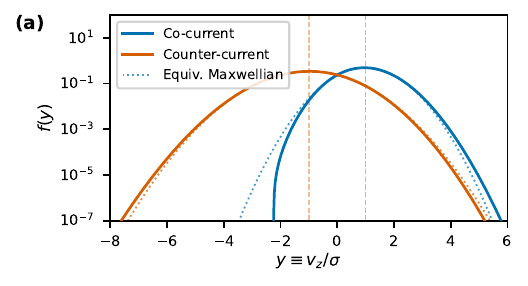}\\[-6pt]
  \includegraphics[width=\columnwidth]{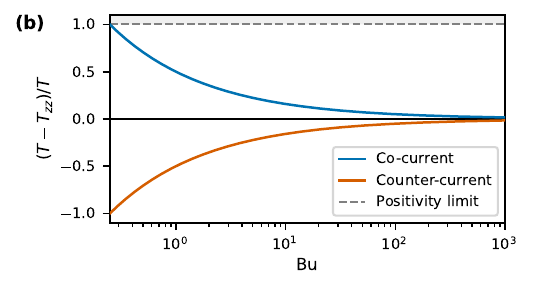}\\[-6pt]
  \includegraphics[width=\columnwidth]{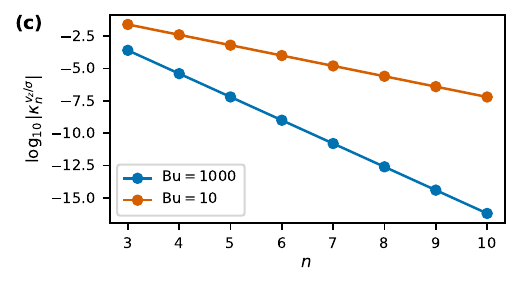}
  \caption{Equilibrium properties of co- and counter- current quadratic
  flows on a Bennett Z pinch at $r = r_p$.
  (a) Distributions and equivalent Maxwellians at $\mu = \pm 1$, $\text{Bu} = 8$.
  (b) Ion temperature anisotropy $(T-T_{zz})/T = \mu/\sqrt{\text{Bu}}$ at $\mu = \pm 0.5$ where
  $\text{Bu}$ is the ion Budker parameter.
  (c) Cumulants $|\kappa_n|$ at $|\mu| = 0.5$, with
  magnetization controlling non-Maxwellianity.}
  \label{fig:bennett-panels}
\end{figure}

The equilibrium distribution at each radius takes the form of a Maxwellian
convolved with a Poisson distribution of jumps of size $\lambda\sigma$
along the flow direction, with rate $|\mu|\,e^{-\lambda a}/\lambda$.
Physically, the distribution is skewed and acquires heavy tails along the flow
direction, with correspondingly depleted tails opposite (see Fig.~\ref{fig:bennett-panels}a).
The cumulants $\kappa_{n\geq 3}$ encode this asymmetry compactly.

The flow direction (co- or counter-current) has distinct consequences
for the existence of equilibrium.

\emph{Co-current flow.}
For $v_z(r) = +V_0\,(r/r_p)^2$ aligned with plasma current
(equivalently $\vec\omega\cdot\vec B < 0$ in the convention of Refs.~\citenum{huba1996, vogman2019, shumlak2020}),
the axial temperature is reduced below thermal ($T_{zz} < T$; see Fig.~\ref{fig:bennett-panels}b).
Requiring temperature positivity $\mu\chi(1 + (r/r_p)^2) < 4$ 
fixes a critical radius,
\begin{equation}
  \frac{r_{\max}^2}{r_p^2} = \frac{4}{\mu\chi} - 1,
  \label{eq:rmax}
\end{equation}
beyond which no equilibrium exists, 
and the equilibrium vanishes entirely when $\mu\chi > 4$.
Co-current sheared-flow equilibrium is permitted only
under sufficient magnetization (small $\chi$, large $\text{Bu}$).
% Equilibrium non-existence for sufficiently high Mach number
% co-current flows is a generic feature of Z-pinch equilibria.
% In a similar way, the co-current linear flux-function equilibrium flow exists
% only up to a certain Mach number / Budker parameter boundary.

\emph{Counter-current flow.}  Taking $v_z(r) = -V_0\,(r/r_p)^2$
opposite the pinch current flips the cumulant signs
and enhances the axial temperature above thermal ($T_{zz} > T$);
variance positivity is therefore automatic and the equilibrium exists at all
radii (see Fig.~\ref{fig:bennett-panels}b).
% The Maxwellian--Poisson convolution can, however, acquire a heavy tail 
% along the counter-current direction, in the low-magnetization regime.

\emph{The role of pinch magnetization.}
The two dimensionless numbers governing the non-Maxwellian statistics
are the Mach number $\mu = V_0/\sigma$ and the
drift parameter $\chi \equiv u_d/\sigma$, with $u_d = j_z/(en)$ the
inter-species axial drift, related to the normalized characteristic
magnetic flux by $\lambda = \chi/4$ (the same $\lambda$ as
Eq.~\eqref{eq:vz-pois}, recast via the Bennett relation
$\mu_0 I^2 = 16\pi N T$), essentially the virial identity
$\text{Bu}\,\chi^2 = 4$, where
$\text{Bu} \equiv \mu_0 q_i^2 N/(4\pi m_i)$ is the ion Budker parameter,
with $N$ the ion line density per unit axial length,
governing ion magnetization.~\cite{crews2025transitional}
Further, the cumulant hierarchy of Eq.~\eqref{eq:cum-pois} decays
with logarithmic slope $\log_{10}[1/(2\sqrt{\text{Bu}})]$.
Therefore, orbit magnetization controls the magnitude of the
non-Maxwellian equilibrium statistics (see Fig.~\ref{fig:bennett-panels}c).
For example, the FuZE device operates at pinch radius $r_p = 0.3$ cm,
with densities $n_e \sim 10^{17}$ cm$^{-3}$ in the long quiescent
period~\cite{shumlak2020} reaching up to $\sim 10^{18}$ cm$^{-3}$
at neutron-emission time,~\cite{goyon2024} yielding a deuterium
Budker parameter $\text{Bu}_d \sim 2$--$20$, well within the
weakly-magnetized regime where non-Maxwellian ion kinetic equilibrium features 
are significant.

\sechead{Equilibrium initialization for kinetic simulation}Initializing
perturbed equilibria is a desired workflow for kinetic simulation,
both particle-based (PIC) and grid-based (continuum).
For the quadratic-flow Bennett pinch above, the mixed
Poisson--Maxwellian statistics admit two complementary approaches.
In continuum codes such as \texttt{WARPM, WARPXM}~\cite{shumlak2011warpm}
and \texttt{Gkeyll},~\cite{mandell2020gkeyll} where the distribution
function is required at each phase-space grid point, the equilibrium
follows directly from evaluating the Gram--Charlier series of
Eq.~\eqref{eq:bell-form}.
For particle codes such as \texttt{WarpX},~\cite{vay2021warpx}
the same equilibrium is sampled by combining a Poisson draw with a
standard-normal draw.
For general flows, the cumulants follow from symbolic expansion of the
flux function (\textit{e.g.}, in \texttt{Mathematica}), and the
equilibrium is sampled via the Cornish--Fisher method,
which produces a target distribution as a polynomial in a
standard-normal sample with cumulant coefficients.~\cite{fisher1960}

\sechead{Conclusions}The electrostatic potential of sheared flow in isothermal kinetic equilibrium,
expressed as a flux function, is the CGF of the distribution function;
equivalently, the plasma density is the MGF.
Linear flux-function flow is therefore the unique admissible polynomial-form
equilibrium, and every other equilibrium sheared flow has non-zero cumulants
to all orders.
The identity applies to Z-pinch, theta-pinch, separable screw-pinch, and planar
Harris equilibria. In the Z pinch, the cumulant decay slope is set by $\rho_L/r_p$,
equivalently the Budker parameter.
The framework yields a direct sampling method for kinetic simulation
initialization. Open directions include dropping radial isothermality to admit a
temperature gradient $\beta = \beta(P_z, P_\theta)$ in the spirit of
Ref.~\citenum{mahajan_li1989}, sheared-flow plasmoid chains from tearing
instability~\cite{tassi2025}, and non-separable screw-pinch equilibria.

\begin{acknowledgments}
The author thanks U. Shumlak and E.T. Meier for careful reading of the manuscript and helpful suggestions,
and N. Reddell and P. Stoltz for stimulating discussions.
\end{acknowledgments}

\section*{Author declarations}

\subsection*{Conflict of interest}
The author has no conflicts to disclose.

\section*{Data availability}

Data sharing is not applicable to this article as no new data were created or analyzed in this study.

\bibliographystyle{aipnum4-2}
\bibliography{refs}

\end{document}